%% file: main.tex
\documentclass[aps,prl,twocolumn,showpacs,superscriptaddress,bibliography]{revtex4-1}
\usepackage{amsbsy,amssymb,amsmath,bm}
\usepackage{graphicx}
\usepackage{braket}
\usepackage{float}
\usepackage{textcomp}
\usepackage{color}
\usepackage[colorlinks=true,linkcolor=red]{hyperref}
\usepackage[sort&compress]{natbib}
\usepackage[normalem]{ulem}
\usepackage[shortlabels]{enumitem}

\usepackage{soul}

\usepackage{amsmath}

\input{epsf}

\begin{document}
\title{
Time molecules with periodically driven interacting qubits
}

\author{K. V. Shulga
}
\affiliation{Center for Emergent Matter Science (CEMS), RIKEN, Wako-shi, Saitama 351-0198, Japan}
\author{I.~Vakulchyk}
\affiliation{Center for Theoretical Physics of Complex Systems, Institute for Basic Science (IBS), Daejeon 34126, Republic of Korea}
\affiliation{Basic Science Program, Korea University of Science and Technology (UST), Daejeon, Korea, 34113}
\author{Y. Nakamura}
\affiliation{Center for Emergent Matter Science (CEMS), RIKEN, Wako-shi, Saitama 351-0198, Japan}
\affiliation{Research Center for Advanced Science and Technology (RCAST), The University of Tokyo, Tokyo 153-8904, Japan}
\author{S. Flach}
\affiliation{Center for Theoretical Physics of Complex Systems, Institute for Basic Science (IBS), Daejeon 34126, Republic of Korea}
\affiliation{Basic Science Program, Korea University of Science and Technology (UST), Daejeon, Korea, 34113}
\author{M. V. Fistul}
\affiliation{Center for Theoretical Physics of Complex Systems, Institute for Basic Science (IBS), Daejeon 34126, Republic of Korea} 
\affiliation{Theoretische Physik III, Ruhr-Universit\"at Bochum, Bochum 44801, Germany}
\affiliation{National University of Science and Technology ``MISIS", Russian Quantum Center, Moscow 119049, Russia}


\date{\today}

\begin{abstract}
We provide numerical evidence 
for a temporal quantum-mechanical interference phenomenon:  time molecules (TM).  
A variety of such stroboscopic states are observed in the dynamics of two interacting qubits subject to a periodic sequence of  
$\pi$-pulses with the period $T$. The TMs appear periodically in time and have a large duration, $\delta t_\mathrm{TM} \gg T$.
All TMs demonstrate an almost zero value of the total polarization and a strong enhancement of the entanglement entropy $S$ up to the maximum value $S=\ln 2$  of a corresponding Bell state.
The TMs are generated by the commensurability of the Floquet eigenvalues and the presence of maximally entangled Floquet eigenstates. The TMs remain stable with detuned system parameters and with an increased number of qubits. The TMs can be observed in microwave experiments with an array of superconducting qubits.
\end{abstract}

\maketitle


\textit{Introduction.} 
The dynamics of a single quantum particle subject to a spatially periodic external potential demonstrates a variety of quantum interference effects~\cite{ficek2005quantum,arndt1999wave,hackermuller2003wave,gerlich2011quantum}. Spatial interference patterns have been observed in the scattering of electrons  on the crystal lattice~\cite{thomson1927diffraction,cowley1995diffraction}, which turned into a routine method to establish the spatial structures of complex molecules and crystals~\cite{zewail20104d,hasselbach2009progress}. 
For quantum systems composed of a few or many interacting quantum particles,
the complex quantum-mechanical dynamics is governed by the interaction and quantum interference effects. It can result in spatially inhomogeneous quantum phases, e.g., quantum vortices and vortex-antivortex pairs  \cite{fisher1990quantum,huang2015quantum},
spin or Bose glasses \cite{edwards1975theory,nohadani2005bose}, etc. These incoherent quantum phases display a broken spatial-translation symmetry of the underlying spatially periodic potential and 
quantum entanglement in space. In particular, spatial quantum entanglement is strongly enhanced as the system is tuned towards a quantum phase transition point~\cite{osterloh2002scaling,amico2008entanglement}.

Temporal quantum interference 
has been observed in the coherent quantum dynamics of a single particle subject to a time-periodic external potential.  Temporal quantum interference effects appear in the form of long-living oscillations of the physical quantities. The frequency of such oscillations is substantially different from the one of the applied time-periodic external potential.  For example, microwave-induced Rabi oscillations have been observed for nuclear spins driven by a resonant ac magnetic field~\cite{rabi1938new}. The frequency of these Rabi oscillations 
is much smaller than the 
frequency of the applied microwave radiation. Also, more complex temporal quantum interference patterns such as Ramsey fringes~\cite{ramsey1950molecular} and spin echo~\cite{hahn1950spin} occur in quantum systems driven by a periodic sequence of dc or ac pulses. Note that these temporal interference patterns have also been observed in various macroscopic artificially prepared two-level qubit systems \cite{nakamura2001rabi,vion2003rabi,bertaina2014rabi}.

An even larger variety of temporal interference patterns can be expected to occur in periodically driven quantum systems composed of many interacting quantum particles~\cite{khemani2016phase,moessner2017equilibration,d2013many,lazarides2015fate,ponte2015many}.  Floquet quantum systems~\cite{grifoni1998driven} have attracted attention recently due to the prediction~\cite{wilczek2012quantum,watanabe2015absence,yao2017discrete}
 of a novel coherent quantum state --- the \textit{time crystal}~\cite{arXiv:1910.10745}.
Subsequent observations~\cite{choi2017observation,zhang2017observation} were reported in
atomic and solid-state systems. Time crystals show stable quantum oscillations with period $2T$, where $T$ is the period of an applied pulse-sequence, in spite of the unavoidable disorder and pulse-sequence imperfection. However, temporal interference patterns and  dynamical properties of quantum states occurring in systems with a few interacting quantum particles have not been studied yet.

In this Letter, we present numerical simulations of the coherent quantum dynamics of periodically driven qubit chains consisting of a few interacting qubits. 
The drive is provided by an externally applied periodic sequence of short spin-flip pulses. We observe various temporal interference patterns in the total polarization of the qubits measured at stroboscopic times. 
These interference patterns depend strongly on two important parameters of the problem ---  the interaction strength between the qubits and the spin-flip pulse imperfection. For a suitable set of these parameters, we obtain dynamical states coined \textit{time molecules} (TM). TMs appear periodically in time, have a long duration, $\delta t_\mathrm{TM} \gg T$,  and  show an almost zero value of the total polarization and a maximal quantum entanglement entropy.
The latter observation indicates that the TMs are in one of the maximally entangled states.  For example for two interacting qubits the observed TMs are in one of the Bell states and demonstrate 
stroboscopic switching between Bell states. The analysis of the Floquet eigenvalues and eigenstates allows us to explain the formation and the dynamical properties of TMs. Such TMs can be directly observed in two-tone dispersive measurements of
short arrays of interacting superconducting qubits.

\textit{Model.} Let us consider a one-dimensional chain of $N$ interacting qubits, i.e., artificially fabricated two-level systems 
subject to a periodic sequence of spin-flip pulses. The 
period of the pulse sequence is $T$. The 
time-dependent Hamiltonian 
acting over a single period $T$ is written in the spin representation as 
\begin{flalign} \label{model_Hamiltonian}
\hat{H}_\mathrm{sys} &=\hat{H}_1 + \hat{H}_2,  \nonumber \\ 
 \hat H_1 &=  \frac{\hbar \alpha}{2 t_1} \sum_{i=1}^N\hat \sigma_{x,i}, ~~0<t<t_1, \nonumber \\
 \hat H_2 &= \frac{\hbar}{T-t_1}\sum_{i=1}^N \delta_i(\hat \sigma_{z,i}+1)  \nonumber \\
 & +\frac{\hbar g}{T-t_1} \sum_{\langle i,j \rangle}(\hat \sigma_{x,i} \hat \sigma_{x,j}+\hat \sigma_{y,i} \hat\sigma_{y,j}),~~t_1<t<T,
\end{flalign}
where $\hat \sigma_{x,i}$, $\hat \sigma_{y,i}$ and $\hat \sigma_{z,i}$ are the corresponding Pauli matrices of the $i$th qubit and $g$ is the dimensionless coupling strength of the exchange interaction between a pair of adjacent qubits, $\langle i,j \rangle$. A single spin-flip pulse is characterized by two parameters, the dimensionless pulse strength $\alpha$ and the pulse duration time $t_1$ ($t_1 \ll T $).
The parameter $\alpha$ is chosen to be close to the $\pi$-pulse: $\alpha=\pi-2\epsilon$. The parameter $\epsilon$ quantifies the imperfection of  the $\pi$-pulse. The implementation of  different coefficients $\delta_i$  in $\hat{H}_2$ allows one to simulate a 
spread of the qubit frequencies. Such model is based on the instantaneous gate approximation as the detuning and interaction between qubits are absent in the time interval, $0<t<t_1$. The validity of such approximation has been discussed in~\cite{appendix}.

The coherent Floquet dynamics of such a system is determined by the discrete time unitary map:
\begin{equation} \label{DynamicsUnitary}
\Psi (nT)=\left[ e^{-i\hat H_2 (T-t_1)/\hbar} \cdot \hat{U}_\epsilon \right]^n \Psi(0),
\end{equation}
where the unitary operator $\hat{U}_\epsilon$ is written explicitly as 
\begin{eqnarray} \label{DynamicsUnitary-2}
    \hat{U}_\epsilon = \bigotimes_i^N  \begin{pmatrix}
        \sin \epsilon   & -i\cos \epsilon \\
        -i\cos \epsilon & \sin \epsilon
    \end{pmatrix}
\end{eqnarray}
Here, $\Psi(0)$ is the wave function of  an initial state, and $n$ is the discrete time measured in units of $T$. 

We present numerical simulations of the above unitary map 
for various arrays composed of a few qubits ($N=2,3,5$). We vary the parameters $\epsilon$ and $g$, and 
measure two observables: the total polarization ($z$-projection of the total spin)
$\langle\hat\sigma_{z}\rangle  =\sum_i^N \langle \hat \sigma_{z,i}\rangle= \bra{\Psi (nT)} \sum_i^N\hat\sigma_{z,i} \ket{\Psi(nT)}$, and the  entanglement entropy $S(nT)$ calculated  by applying standard methods of quantum statistical physics~\cite{eisert2010colloquium,appendix}.

\begin{widetext}


\begin{figure}
\centering
\includegraphics[width=7.2in,angle=0]{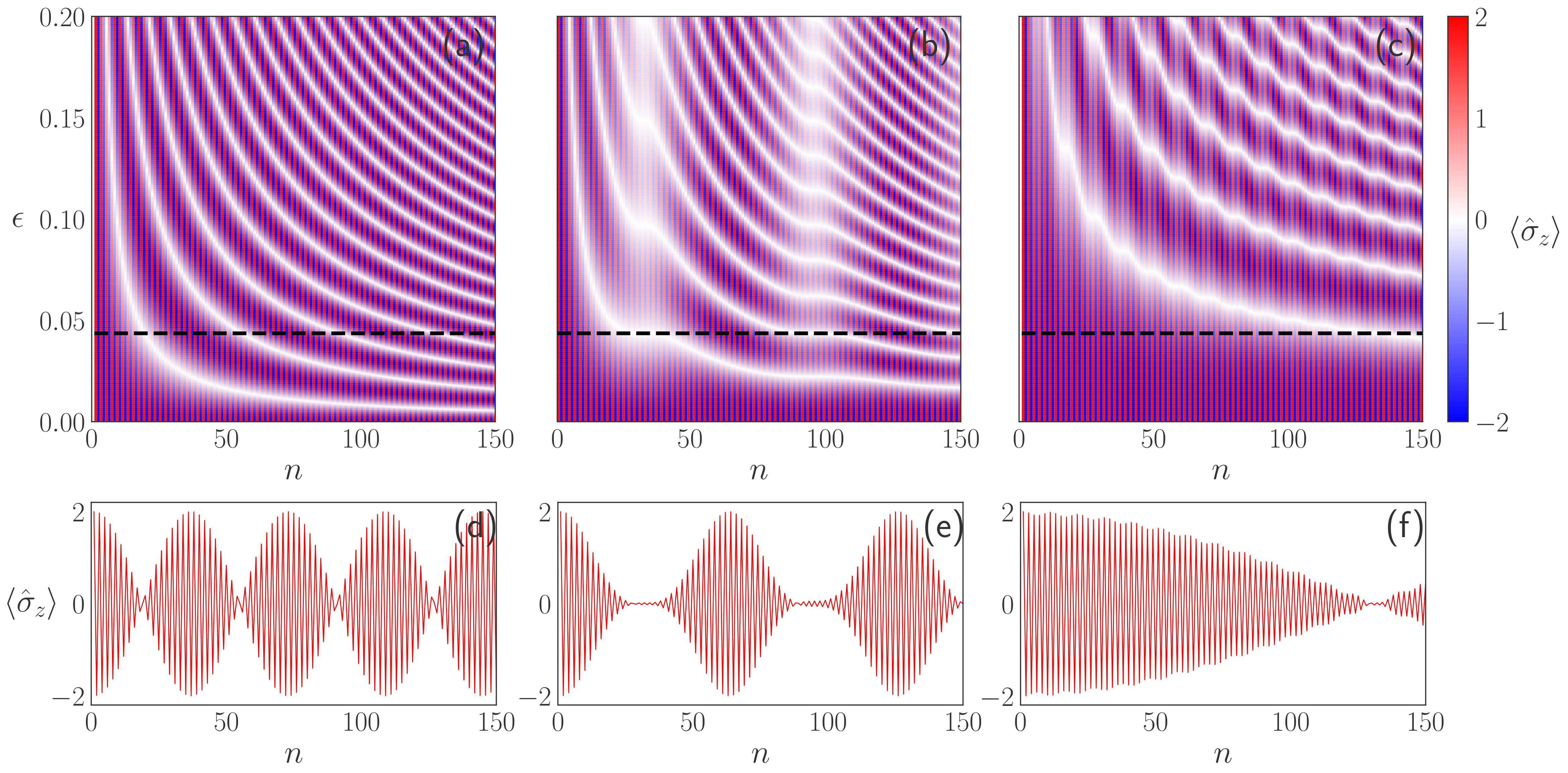}
\caption{ 
(a)--(c) Instantaneous total polarization $\langle \hat \sigma_z \rangle$ 
as a function of  the discrete time $n$ 
and the pulse imperfection $\epsilon$ for two interacting qubits
. The interaction 
strengths are (a)~$g=0$, (b) $g=0.05$, and (c)  $g=0.3$, respectively. The detuning $\delta_{1,2}$ between the qubits is set to zero. 
(d)--(f) Cross-sections of the plots in (a)--(c) at $\epsilon=0.0436$ indicated with dashed lines.}
\label{Pic.1}
\end{figure}

\end{widetext}
\begin{figure}[h!]
\centering
\includegraphics[width=3.0in,angle=0]{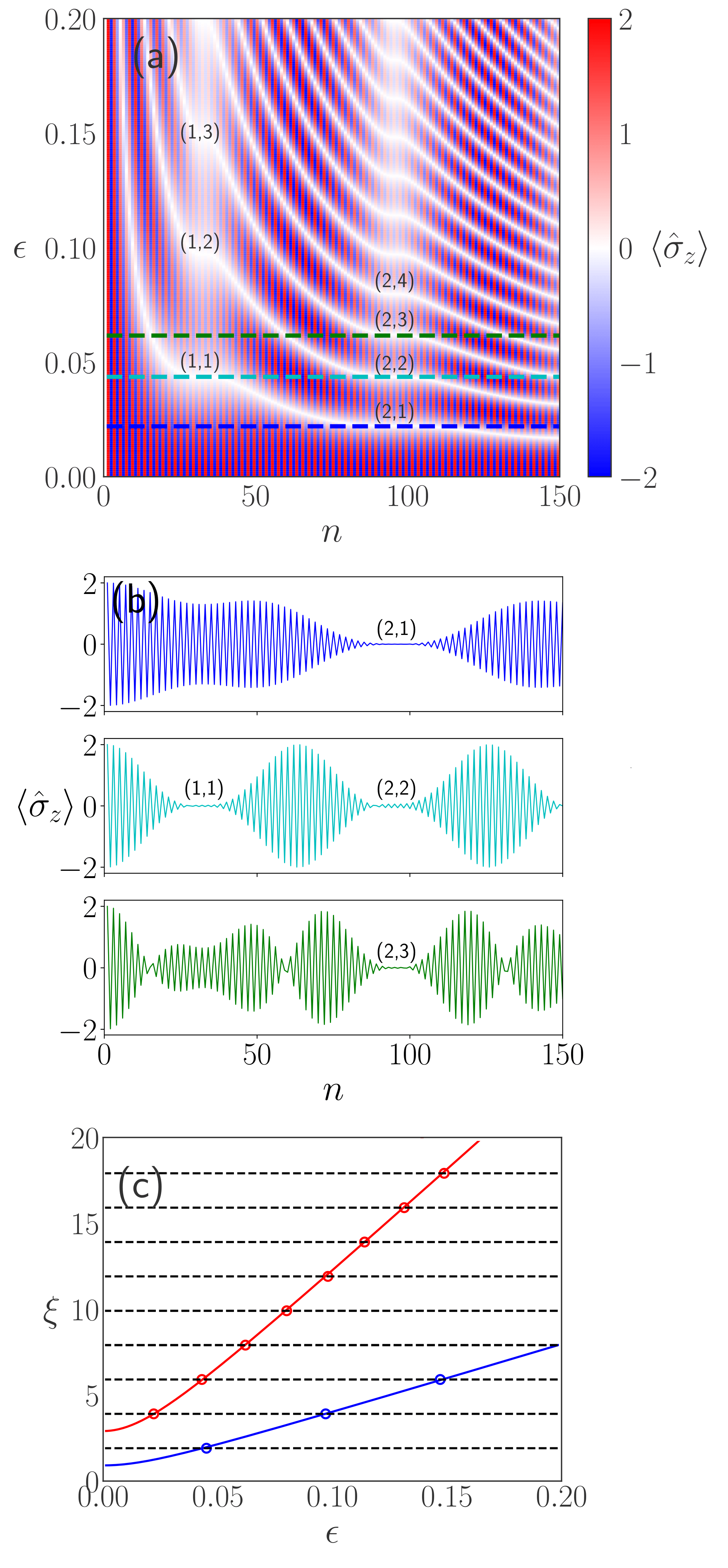}
\caption{
(a)~Instantaneous total polarization $\langle \hat{\sigma_z} \rangle$ as a function of the discrete time $n$ and the pulse imperfection $\epsilon$ for two interacting qubits with $g=0.05$.
The 
indices $(k,l)$ indicate different TM states. 
(b)~Cross-sections at the values of $\epsilon$ shown in~(a) with dashed lines. 
(c)~Generalized ratio of the quasienergies, $\xi_k=(k+1)(\pi-\epsilon_\mathrm{FL})/g$, as a function of $\epsilon$ for different values of commensurability parameter $k=1$ (blue curves) and $k=2$ (red curves). The TMs of the first (second) group are 
observed when $\xi_k$ with $k=1$ ($k=2$) becomes an even integer (circles at the crossing points with dashed lines).  }
\label{Pic.2}
\end{figure}
\begin{figure}[h!]
\centering
\includegraphics[width=3.0in,angle=0]{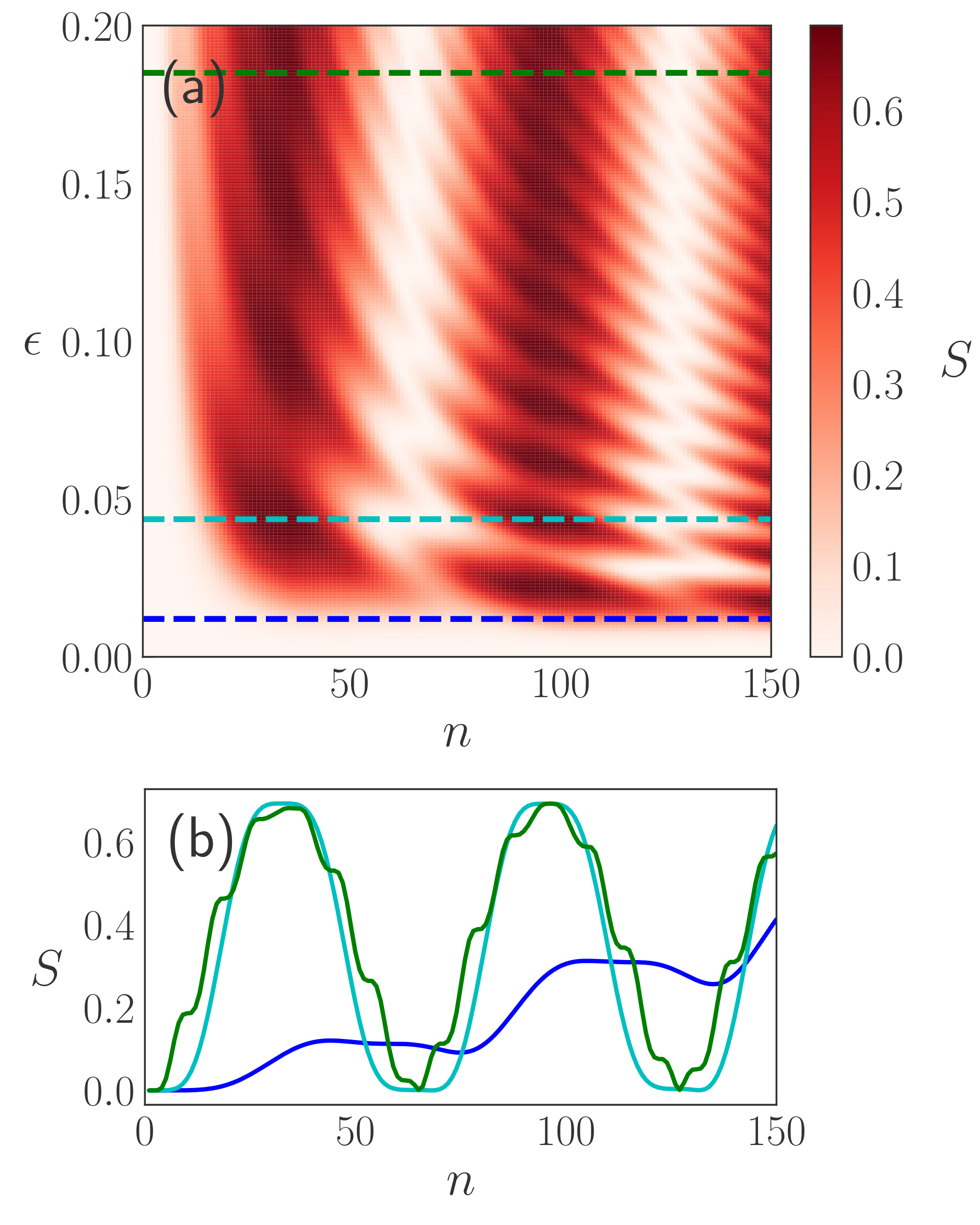}
\caption{
(a)~Entanglement entropy $S$ as a function of the discrete time $n$ and the pulse imperfection $\epsilon$ for two interacting qubits~($g=0.05$).
(b)~Cross-section of the plot in (a) at $\epsilon=0.012$ (blue), $\epsilon=0.0436$ (cyan), and $\epsilon=0.185$ (green), respectively. The maximum entropy $S=\ln 2$ is obtained at times where TMs are formed~(Fig.~\ref{Pic.2}).
}
\label{Pic.3}
\end{figure}
\begin{figure}[h!]
\centering
\includegraphics[width=3.0in,angle=0]{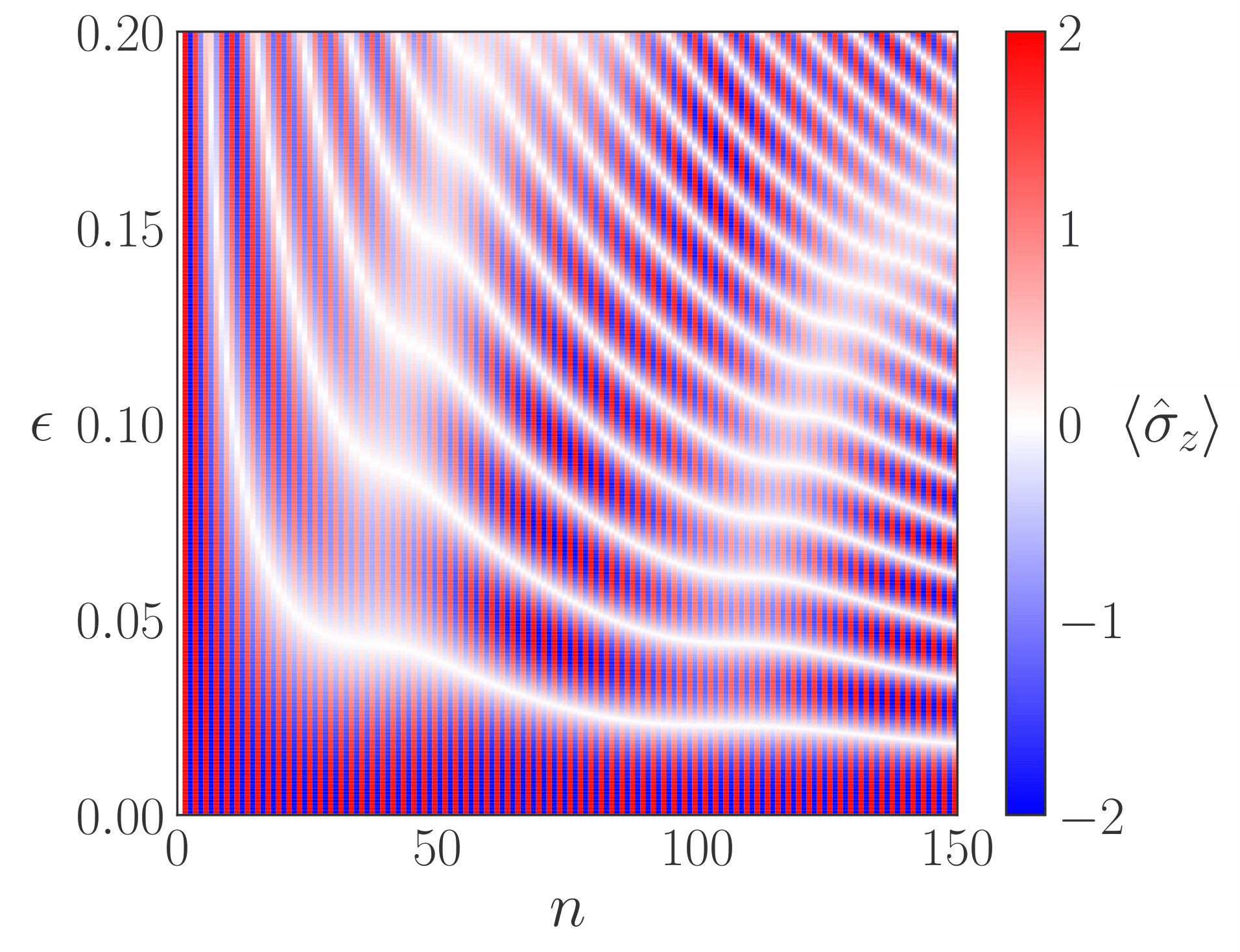}
\caption{Instantaneous total polarization $\langle \hat{\sigma_z} \rangle$ as a function of the discrete time $n$ and the pulse imperfection $\epsilon$ for two interacting qubits in the presence of a large detuning, \textbf{$\delta_{1}=0$} and \textbf{$\delta_{2}=0.7$.}
The qubit interaction strength is set to $g=0.05$. 
}
\label{Pic.4}
\end{figure}

In the numerical simulations, the initial ferromagnetic state is chosen as $\ket{\uparrow \uparrow...\uparrow }$. A small number of qubits and the periodic time-dependence of the Hamiltonian $\hat{H}_\mathrm{sys}$ [Eq.~(\ref{model_Hamiltonian})] allows to study the dynamics of the system up to large stroboscopic times, e.g., $n >1000$. The typical dynamics for times $0<n<150$ is shown in Figs.~\ref{Pic.1}--\ref{Pic.4}.   The pulse-imperfection parameter $\epsilon$ is varied from $0$ to $0.2$. 

\textit{Total polarization.} We first focus on the dynamics of two identical interacting qubits with $\delta_{1,2} \equiv \delta_1 - \delta_2 = 0$. 
For all values of interaction strength $g$, the dynamics shows stroboscopic oscillations with the frequency, $\pi/T$. 
For $g=0$, the total polarization displays quantum beats with a large characteristic period $  \pi T/(2\epsilon)  \gg T$ (see the detailed analysis of quasienergies in  \cite{appendix}). The quantum beats are shown in the plots of  Figs.~\ref{Pic.1}(a) and (d).
Notice that at discrete times $n\approx\pi (2m+1)/(4\epsilon)$, $m=0,1,2...$, the total polarization is zero [white lines in  Fig.~\ref{Pic.1}(a)]. A slight increase of the interaction strength $g \ll \epsilon$ results in a distortion of the quantum oscillations, and a moderate increase of the beating period (not shown). 
In the opposite case $g \gg \epsilon$ the quantum beats in Figs.~\ref{Pic.1}(c) and (f) gain very large periods $ \pi g T/(2\epsilon^2)$ (see the analysis in~\cite{appendix}).

A most interesting and intriguing dynamics is obtained in the range of intermediate values of $g \simeq \epsilon$ [Figs.~\ref{Pic.1}(b) and \ref{Pic.1}(e)]. For particular values of $\epsilon$ one can see a proliferating set of periodically distributed \textrm{flat regions} in the dependence of the total polarization on time. 
In these flat regions $\langle \hat \sigma_z  \rangle$ displays almost vanishing  fluctuations around zero. The time duration of the flat regions $\delta t_\mathrm{TM}$ varies from $\simeq 10 T$ up to $ \simeq 30 T$ [Figs.~\ref{Pic.2}(a) and (b)].
We coin these long-lived metastable states of the Floquet dynamics of two interacting qubits as Time Molecule (TM) states.

A detailed study of the TM dynamics results in an interesting observation: for a fixed value of  $g$ these flat regions are grouping around particular times
which are periodically distributed.  For example, for $g=0.05$ and the time interval of $0<n<150$ we obtain two groups of flat regions  of $\langle \hat \sigma_z  \rangle$ concentrating on times around $n=30$ and $n=100$ [Fig.~\ref{Pic.2}(a)]. 
We label the flat regions with indices $(k,l)$ as shown in  Fig.~\ref{Pic.2}(a) for $k=1$ and $~2$. 
The  dynamics of the TMs 
of different groups 
is presented in Fig.~\ref{Pic.2}(b).


\textit{Entanglement entropy.} 
Here, we follow the time dependence of entanglement entropy $S(nT)$ for different values of $\epsilon$ and $g$. 
Typical results are presented in the color plot of Fig.~\ref{Pic.3}(a) for the interaction strength $g=0.05$. The entanglement entropy demonstrates a periodic dependence on time. 
Since the initial state in our simulations was chosen to be a product state, we observe that at discrete times $n< 1/\epsilon$, 
the entanglement entropy takes small values (the left-most white regions). However, one can see that in the flat regions of $\langle \hat \sigma_z  \rangle$, where the TMs are formed, the entanglement entropy reaches its maximum value of $S=\ln 2$.

We observe various types of the time-dependence of $S(nT)$ (see Fig.~\ref{Pic.3}(b)) which depend on the ratio of the interaction strength $g$ and the pulse imperfection $\epsilon$. For $g \gg \epsilon$  a slight increase of $S$ up to some moderate value of $S \simeq 0.3$ is observed in the time region $0<n<150$ (blue line).
For $g \ll \epsilon$, $S(nT)$ shows an additional and frequent step-like incremental and decremental modulation (green  curve). The positions of the steps correspond to the stroboscopic times where the total polarization vanishes~[see Fig.~\ref{Pic.2}(a)]. 
For $g \simeq \epsilon$ regime, $S(nT)$  performs strongly anharmonic and sharp modulations between values 0 to $\ln 2$~(or vice versa) at the stroboscopic times where the TMs start to form (or disappear). 
These observations indicate that the metastable TMs  are in a maximally entangled state, and the value of $S=\ln 2$ does not vary during the TM duration time $\delta t_\mathrm{TM}$.

\textit{Discussion.} To explain the flat regions in the time dependence of $\langle \hat \sigma_z  \rangle$ and the enhancement of the entanglement entropy $S(nT)$,
we calculate the quasienergies and Floquet eigenstates of the periodically driven system~\cite{grifoni1998driven,appendix}.
In the general case of two interacting periodically driven qubits, there are four quasienergies varying with $\epsilon$ and $g$. However, for this particular Floquet problem 
the quasienergies are fixed at $E_1 = 0$; $E_2 = 2 \hbar g/T$; $E_3= \hbar \epsilon_\mathrm{FL}/T$;  $E_4 = -\hbar (\epsilon_\mathrm{FL} + 2g)/T$, where a   dimensionless parameter $\epsilon_\mathrm{FL}=\pi -g -\sqrt{g^2+4\epsilon^2}$ as $\{\epsilon, g\} \ll 1$ (see details in the Supplementary Material~\cite{appendix}).
The Floquet dynamics shows oscillations with a few small frequencies, i.e.\ $\omega_1= \pi/T-E_3/\hbar=(\pi-\epsilon_\mathrm{FL})/T$ and $\omega_2=E_2/\hbar=2g/T$, and we observe that the TMs form as the generalized ratio of these frequencies  $\xi_{k}(\epsilon,g)=2(k+1)\omega_1/\omega_2=(k+1)(\pi-\epsilon_\mathrm{FL})/g$ is equal to $\ell$, where $k$ is the commensurability parameter, and $\ell$ is an even integer number.  In Fig.~\ref{Pic.2}(c) the dependencies of $\xi_{k}(\epsilon,g)$ for two groups of TMs with $k=1$ and $k=2$ are presented. 

The largest entanglement entropy of the TMs is explained by inspecting the properties of the Floquet eigenstates. We find two  Floquet eigenstates with maximal entanglement entropy $S=\ln 2$: $\psi_1=(1/\sqrt{2})(\ket{\uparrow,\uparrow}-\ket{\downarrow,\downarrow})$ and  $\psi_2=(1/\sqrt{2})(\ket{\uparrow,\downarrow}-\ket{\downarrow,\uparrow})$. These two Floquet eigenstates form a pair of Bell states. The two remaining Floquet eigenstates  $\psi_{3,4}$ have low values of $S$ which in addition strongly depend on  $\epsilon$ and $g$. 
The eigenstates $\psi_1$ and $\psi_{3,4}$ are symmetric with respect to a permutation of the qubits, while $\psi_2$ is anti-symmetric. Since the initial state is symmetric and the Hamiltonian in Eq.~(\ref{model_Hamiltonian}) for identical qubits conserves the state symmetry,  the eigenstate $\psi_2$ is not excited during the observed dynamics. Another important property of the Floquet eigenvectors is that the total polarization is zero in all the Floquet eigenstates. Moreover, the Floquet eigenstate $\psi_1$ does not vary with time ($E_1$ is zero). Thus, we arrive at the following scenario of the two-qubit dynamics: for finite $\epsilon$ and $g$ the system is oscillating with period $T$ between different states and slowly approaches the states in which  $\langle \hat \sigma_z \rangle=0$  [white lines in Figs.~\ref{Pic.1} and \ref{Pic.2}]. 
After that, the system arrives at the TM state where
the dynamics is a stroboscopic switching between the two maximally
entangled Bell states: $(1/\sqrt{2})(\ket{\uparrow,\uparrow}-i\ket{\downarrow,\downarrow})$ and  $(1/\sqrt{2})(\ket{\uparrow,\uparrow}+i\ket{\downarrow,\downarrow})$, and a rather slow leakage into other eigenstates. 
These maximally entangled Bell states are formed from the particular
superposition of eigenstates $\psi_1$, $\psi_3 \exp[-iE_3 n T/\hbar]$
and $\psi_4 \exp[-iE_4 n T/\hbar]$.

So far, our numerical study was carried out for identical qubits $\delta_1 = \delta_2 = 0$. For the case of different qubit frequencies with $\delta_1=0$ and $\delta_2=0.7$, the dynamics of the total polarization
is presented in Fig.~\ref{Pic.4}. The TMs of both groups are still present, but the stroboscopic formation times of the TMs  shift to larger values. Such shifts are especially pronounced for large values of $\epsilon$
[cf.~Figs.~\ref{Pic.2}(a) and \ref{Pic.4}]. 
Also we verify (see the Supplementary material~\cite{appendix}) that slightly different spin-flips pulses with unequal values of $\epsilon_{1}$ and $\epsilon_{2}$ lead to shifts of TMs formation times as $\epsilon_\mathrm{add}=|\epsilon_1-\epsilon_2| \ge g$.

We have also carried out numerical calculations of qubits chains with a larger number of qubits, i.e., $N=3$ and 
$N=5$. 
The region of $\epsilon$ where the flat regions which indicate the appearance of the TMs completely disappear, and stable oscillations with the  period $2T$  are observed, is greatly extended with an increased number of qubits (compare Fig.~\ref{Pic.2}a and Fig.~S2 in the Supplementary Material~\cite{appendix}). These oscillations indicate the appearance of the time-ordering in a system with a few interacting qubits. 
For larger values of $\epsilon$ the flat regions are shifted and distorted forming  complex patterns in time-space domain (see Fig.~S2 in the Supplementary Material~\cite{appendix}).
Such complex patterns can be considered as the precursors of temporal and spatial  many-body localized 
states arising in periodically driven  systems with a large number of interacting quantum particles \cite{d2013many,lazarides2015fate,ponte2015many,yao2017discrete}. 

The TM states can be  directly observed in  microwave experiments with an array of interacting superconducting transmon qubits. In such a setup the periodic sequence of ac pulses is applied to each qubit. 
The frequency of ac pulses $\omega$  has to be chosen close to the qubit frequencies $\omega_i$ and the amplitude of the pulses determines the amplitude $\alpha$ in Eq.~(\ref{model_Hamiltonian}). The spread of parameters reads $\delta_i=\omega-\omega_i$. The  qubit population imbalance (total polarization) $\langle \hat \sigma_z \rangle$ can be directly measured with a dispersive readout regularly used for the study of quantum dynamics of superconducting qubit networks \cite{blais2007quantum,majer2007coupling,jerger2012frequency}.

\textbf{Acknowledgements}
This work was partly supported by the Institute for Basic Science, Project Code (IBS-R024-D1), and JST ERATO (No.~JPMJER1601). M. V. F. 
thanks the partial financial support of the Russian Science Foundation, Project (19-42-04137). 
We are grateful to Franco Nori for the name of  ``time molecule" and thank Alexander Cherny for valuable discussions. 

\input{main.bbl}

\end{document}


\title{
Supplementary material for \\
``Time molecules with periodically driven interacting qubits''}

\author{K. V. Shulga
}
\affiliation{Center for Emergent Matter Science (CEMS), RIKEN, Wako-shi, Saitama 351-0198, Japan}
\author{I.~Vakulchyk}\affiliation{Center for Theoretical Physics of Complex Systems, Institute for Basic Science (IBS), Daejeon 34126, Republic of Korea}
\affiliation{Basic Science Program, Korea University of Science and Technology (UST), Daejeon, Korea, 34113}
\author{Y. Nakamura}
\affiliation{Center for Emergent Matter Science (CEMS), RIKEN, Wako-shi, Saitama 351-0198, Japan}
\affiliation{Research Center for Advanced Science and Technology (RCAST), University of Tokyo, Japan}
\author{S. Flach}
\affiliation{Center for Theoretical Physics of Complex Systems, Institute for Basic Science (IBS), Daejeon 34126, Republic of Korea}
\affiliation{Basic Science Program, Korea University of Science and Technology (UST), Daejeon, Korea, 34113}
\author{M. V. Fistul}
\affiliation{Center for Theoretical Physics of Complex Systems, Institute for Basic Science (IBS), Daejeon 34126, Republic of Korea}
\affiliation{Theoretische Physik III, Ruhr-Universit\"at Bochum, Bochum 44801, Germany}
\affiliation{National University of Science and Technology ``MISIS", Russian Quantum Center, Moscow 119049, Russia}

\date{\today}
\maketitle

\section{I. Entanglement entropy of two interacting qubits}
The time dependence of the entanglement entropy $S(t)$ for the coherent Floquet dynamics of two interacting qubits  is calculated by the following method \cite{eisert2010colloquium,dirac1930note}. At stroboscopic times $nT$ the wave function $\Psi(nT)$ is expressed as
\begin{equation} \label{StrobWavefunction}
\Psi(nT)=\sum_{i=1}^4 C_i(nT)\psi_i,
\end{equation}
where the basis wave functions are $\psi_1=\ket{\uparrow \uparrow}$, $\psi_2=\ket{\uparrow \downarrow}$, $\psi_3=\ket{\downarrow \uparrow}$ and $\psi_4=\ket{\downarrow \downarrow}$. The full density matrix $\rho$ with the dimension of $4$ is constructed as following:
\begin{equation} \label{DensityMatrix}
(\rho(nT))_{ij}=C^\ast_{i}(nT)  C_{j}(nT) .
\end{equation}

Next, we perform a partial trace operation, obtaining  the $2 \times 2$ reduced density matrix $\tilde \rho(nT)$ written explicitly as
\begin{eqnarray} \label{matrix_r}
    \tilde \rho (nT)=  \begin{pmatrix}
        \tilde \rho_{11}=\rho_{1,1} + \rho_{2,2} ;&   \tilde \rho_{12}=\rho_{1,3} + \rho_{2,4}\\
          \tilde \rho_{21}=\rho_{3,1} + \rho_{4,2}; &   \tilde \rho_{22}=\rho_{3,3} + \rho_{4,4}
    \end{pmatrix}.
\end{eqnarray}
Calculating the eigenvalues $\alpha_{1,2}(nT)$ of the reduced density matrix $\tilde \rho(nT)$ we obtain the entanglement entropy
\begin{equation}
S(nT) = -\sum_{i=1}^2 |\alpha_i| \log(|\alpha_i|)
\end{equation}
Note that $S(nT)=0$ in the absence of quantum entanglement [i.e. the wave function $\Psi(nT)$ becomes a separable product state]. At variance,  for completely entangled states (Bell states) the entanglement entropy reaches its maximum value  $S(nT)=\ln 2$.


\section{II. The quasienergies and the Floquet eigenstates of two-interacting qubits}
Here, we calculate the quasienergies of two identical interacting qubits subject to the time-periodic pulse sequence. The coherent quantum dynamics of such system is controlled  by the time-periodic Hamiltonian:
\begin{flalign} \label{model_Hamiltonian}
\hat{H}_\mathrm{sys} &=\hat{H}_1 + \hat{H}_2,  \nonumber \\
 \hat H_1 &=  \frac{\hbar \alpha}{2t_1} (\hat \sigma_{x,1}+\hat \sigma_{x,2}), ~~0<t<t_1, \nonumber \\
 \hat H_2 &=   \frac{\hbar g}{T-t_1}(\hat \sigma_{x,1} \hat \sigma_{x,2}+\hat \sigma_{y,1} \hat\sigma_{y,2}),~~t_1<t<T.
\end{flalign}
The pulse parameter $\alpha$ is chosen to be close to the $\pi$ pulse:
$\alpha =\pi-2\epsilon$, and $t_1 \ll T$. In this particular case the dynamics is determined by the discrete time unitary map as follows:
\begin{equation} \label{DynamicsUnitary}
\hat \Psi (T)=\hat F \hat \Psi(0)=\left[ \hat{U}_g \cdot \hat{U}_\epsilon \right] \hat \Psi(0),
\end{equation}
where the unitary operator of $U_\epsilon$ is written explicitly as
\begin{eqnarray} \label{DynamicsUnitary-2}
    \hat{U}_\epsilon =  \begin{pmatrix}
        \sin \epsilon   & -i\cos \epsilon \\
        -i\cos \epsilon & \sin \epsilon
    \end{pmatrix} \otimes
    \begin{pmatrix}
        \sin \epsilon   & -i\cos \epsilon \\
        -i\cos \epsilon & \sin \epsilon
    \end{pmatrix}
\end{eqnarray}
and the unitary interaction operator of $U_g$ is written as
\begin{eqnarray} \label{DynamicsUnitary-G}
    \hat{U}_g =  \begin{pmatrix}
        1   & 0 & 0 & 0 \\
       0 & \cos (2g) & i \sin (2g) & 0 \\
       0 & i \sin (2g) &  \cos (2g) & 0 \\
       0 & 0 & 0 & 1
    \end{pmatrix}
\end{eqnarray}
The unitary map $\hat F$ has eigenvalues of the form, $\exp(-i E_i T/\hbar)$, where $E_i$ are quasienergies, and $i=1,..4$.

Next, we explicitly calculate the quasienergies $E_i$ in the particular limit as $\{\epsilon,g \} \ll 1$. In this limit the matrix $\hat F$ is written as:
\begin{eqnarray} \label{DynamicsUnitary-F}
    \hat{F} =  \begin{pmatrix}
        0   & -i\epsilon & -i\epsilon & -1 \\
       -i\epsilon & -2i g & -1& -i\epsilon \\
       -i\epsilon &-1 & -2i g & -i\epsilon\\
       -1 & -i\epsilon & -i\epsilon & 0
    \end{pmatrix}
\end{eqnarray}
Diagonalizing the matrix $\hat F$ we obtain the quasienergies $E_i$ as: $E_1 = 0$; $E_2 = 2\hbar g/T$; $E_3=\hbar \epsilon_\mathrm{FL}/T$;  $E_4 = -\hbar (\epsilon_\mathrm{FL} + 2g)/T$, where  $\epsilon_\mathrm{FL}=\pi -g -\sqrt{g^2+4\epsilon^2}$.
Thus, one can obtain the quasienergy $\epsilon_\mathrm{FL}=\pi-2\epsilon$ for $g \ll \epsilon$ and $\epsilon_\mathrm{FL} =\pi-2g-2\epsilon^2/g$ for the opposite regime $g \gg \epsilon$.

The corresponding Floquet eigenvectors are written in an explicit form as:
\begin{flalign} \label{FloquetEigenvectors}
\psi_1=(1/\sqrt{2})(\ket{\uparrow,\uparrow}-\ket{\downarrow,\downarrow}), ~~~~~~~~~~~~~~~~~~~~~~~~~~~~~~~~~~~~\nonumber \\
\psi_2=(1/\sqrt{2})(\ket{\uparrow,\downarrow}-\ket{\downarrow,\uparrow}),  ~~~~~~~~~~~~~~~~~~~~~~~~~~~~~~~~~~~~\nonumber \\
\psi_3=(\beta/\sqrt{2(1+\beta^2)})[\ket{\uparrow,\uparrow}+\ket{\downarrow,\downarrow}-(1/\beta) (\ket{\uparrow,\downarrow}+\ket{\downarrow,\uparrow})] \nonumber \\
\psi_4=(1/\sqrt{2(1+\beta^2)})[\ket{\uparrow,\uparrow}+\ket{\downarrow,\downarrow}+\beta (\ket{\uparrow,\downarrow}+\ket{\downarrow,\uparrow})], \nonumber \\
\end{flalign}
where $\beta=(g+\sqrt{g^2+4\epsilon^2})/(2\epsilon)$.

The Floquet dynamics displays quantum beats with various characteristic frequencies, e.g. in the limit of $g=0$ the quantum beats with the frequency $2(\pi-\epsilon_\mathrm{FL})/T$ are obtained, and as $g \gg \epsilon$ the quantum beats with an extremely small frequency $2(\pi-\epsilon_\mathrm{FL}-2g)/T$ are obtained.

\section{III. Robustness of the Floquet dynamics for various pulse imperfections}

We also study the variation of the Floquet dynamics of two interacting qubits in the presence of two slightly different spin-flip pulses with not equal values of $\epsilon_{1}$ and $\epsilon_{2}$. We observe that the influence of additional $\epsilon_\mathrm{add}=\epsilon_1-\epsilon_2$ is similar to other types of disorder, e.g. $\delta_i$. The Floquet dynamics does not change for $g \ge \epsilon_\mathrm{add}$ [Fig.~$S1$(a)] and shows a great distortion in the opposite limit of $g \ll \epsilon_\mathrm{add}$~[Fig.~$S1$(b)].

\begin{figure}[h!]
\centering
\includegraphics[width=3.0in,angle=0]{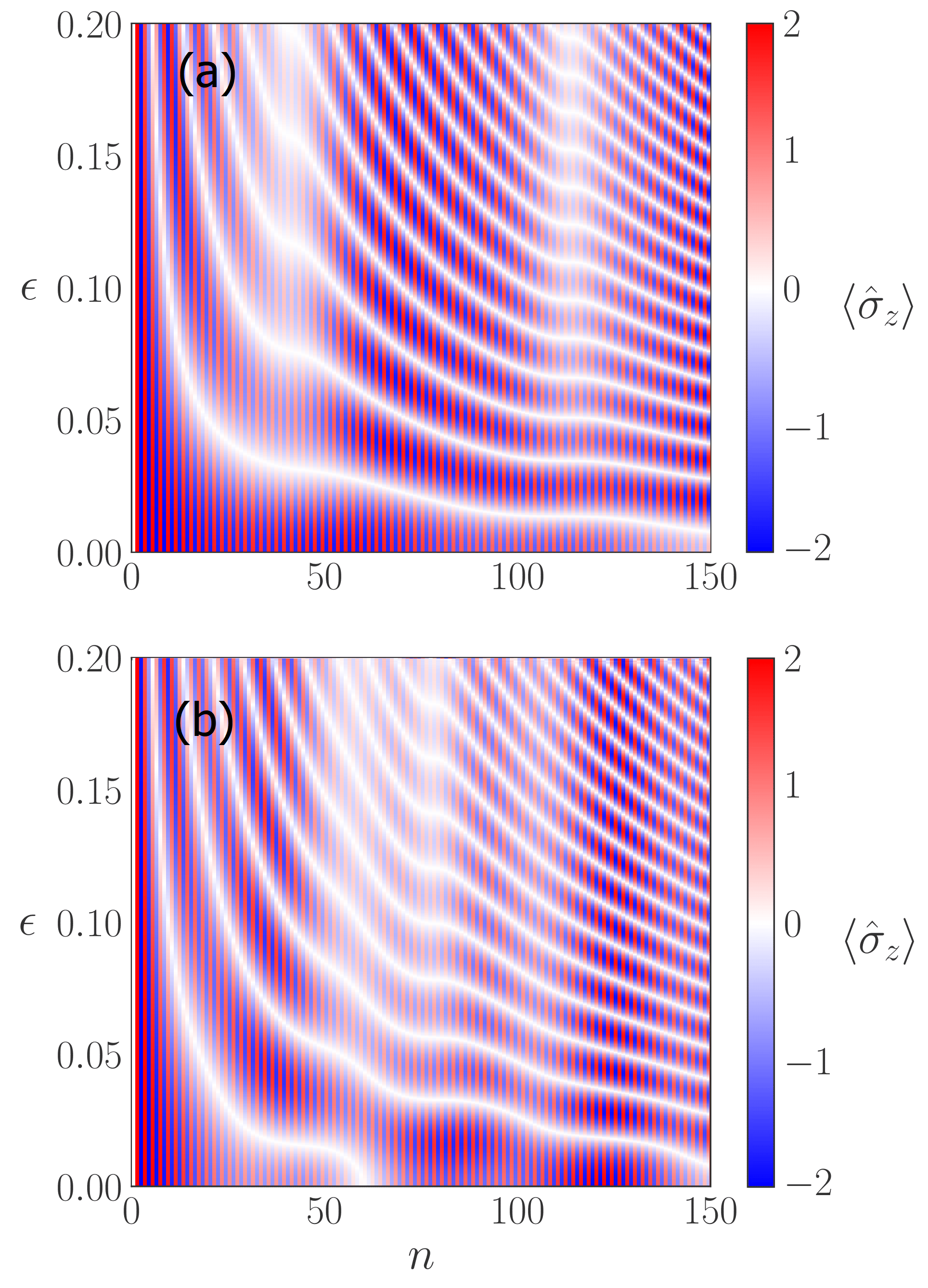}
\caption{ Instantaneous total polarization $\langle \hat{\sigma_z} \rangle$ as a function of the discrete time $n$ and the pulse imperfection $\epsilon$ for two interacting qubits in the presence of nonuniform pulse imperfections, (a)~$\epsilon_\mathrm{add}=0.03$ and (b)~$\epsilon_\mathrm{add}=0.06$.
The disorder in qubits frequencies is absent, and the interaction strength is set to $g=0.05$.}
\label{Pic.6}
\end{figure}

\section{IV. Floquet dynamics of a large number of interacting qubits}

We have also carried out numerical calculations of qubits chains with a larger number of qubits, i.e.\ for $N=3$ and
$N=5$. The results for the instantaneous total polarization are presented in Fig.~S2.

\begin{figure}[h!]
\centering
\includegraphics[width=3.0in,angle=0]{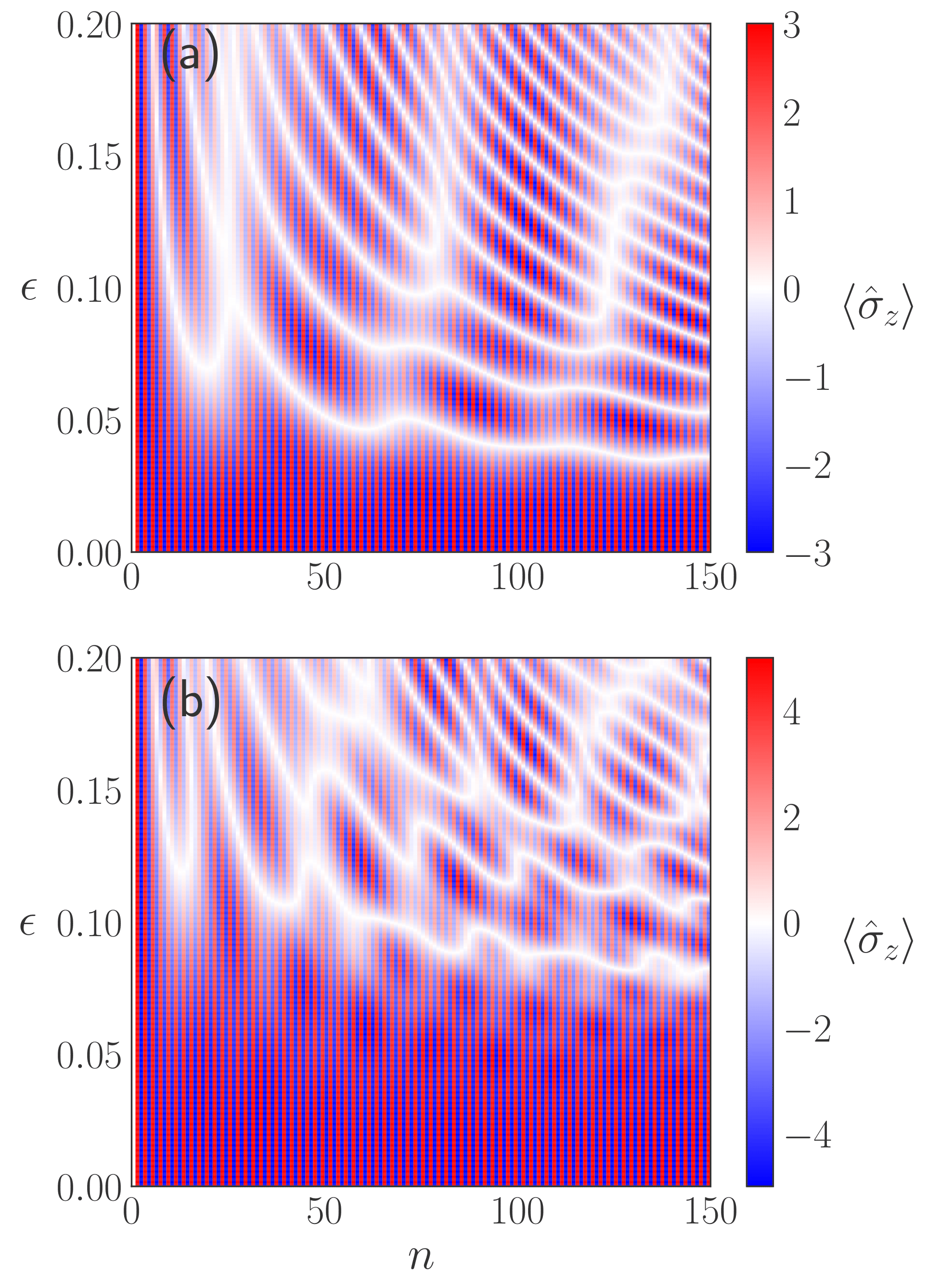}
\caption{
Instantaneous total polarization $\langle \hat{\sigma_z} \rangle$ as a function of the discrete time $n$ and the pulse imperfection $\epsilon$ for $N$ interacting qubits, (a)~$N=3$ and (b)~$N=5$.
The disorder in qubits frequencies is absent, and the interaction strength is set to $g=0.05$.}
\label{Pic.5}
\end{figure}

\section{V. Evaluation of the instantaneous gate approximation for experiments}

In the main text, we use an instantaneous gate approximation in which the Hamiltonian $H_1$ acts during the time, $t_1 \ll T$. This model means that the interaction between qubits is absent (turned off) during the time $t_1$. 
Since this approximation is rather difficult to satisfy in experiments we study the influence of this approximation on the Floquet dynamics of the TM. A simple evaluation of additional inhomogeneity that the system will receive due to the presence of additional interaction during the spin-flip pulse is $\pi*gt_1/T$. This imperfection can be viewed as an additional form of disorder that our system can withstand.

To check that, we introduce the time $t_1$ explicitly in the computational model, and for the ratio $T/t_1 \geq 10$ we still observe a stable dynamics of the TMs. Under this condition, the influence of the presence of the interaction $g$ during the $t_1$ time is negligible. Therefore, the TMs can be observed in experiments e.g.\ with superconducting qubits using current experimental methods.
\input{sup.bbl}

%% file: main.bbl
%

%% file: sup.bbl
%